\documentclass[12pt]{article}%
\usepackage{amssymb}
\usepackage{amsmath}
\usepackage{amsfonts}
\usepackage{cite}
\usepackage{graphicx}
\usepackage{subfig}%

\begin{document}

\author{I.P. Smirnov\\Institute of Applied Physics RAS, \\46 Ul'yanova Street, Nizhny Novgorod, Russia}
\title{Boundary conditions for matrix of variations in refractive waveguides with
rough bottom}
\date{}
\maketitle

\begin{abstract}
The solution of the Helmholtz equation in optical semiclassic approximation is
associated with the calculation of ray paths and matrices of variations. The
transformation rules for elements of matrices on the boundaries of the
waveguide are obtained.

\end{abstract}

\textbf{Key words:} geometry acoustics, matrix of variations, boundary conditions

Consider a two-dimensional underwater acoustic waveguide with the sound speed
$c=c\left(  r,z\right)  $ being a function of depth, $z$, and range, $r$, and
curve bottom $z=z_{b}\left(  r\right)  $.

It is well-known  that in optical semiclassical approximation the sound wave
field can be expressed through parameters of ray trajectories governed by the
Hamilton equation \cite{Viro}%

\begin{equation}
\left\{
\begin{array}
[c]{l}%
\frac{d}{dr}z=\frac{\partial H}{\partial p},\\
\frac{d}{dr}p=-\frac{\partial H}{\partial z},
\end{array}
\right.  \label{Luch}%
\end{equation}
with the Hamilton function
\[
H=-\sqrt{n^{2}-p^{2}}.
\]
The variable $p=n\sin\vartheta$ is the ray pulse, $\vartheta$ is ray grazing
angle, $n=n\left(  r,z\right)  =c_{0}/c\left(  r,z\right)  $ is the refractive
index, where $c_{0}$ is a reference sound speed.

To find a ray trajectory starting from the source point $\left(  r_{0}%
,z_{0}\right)  $ at the angle $\vartheta_{0}$ we must solve the system
(\ref{Luch}) with the initial conditions $z=z_{0},\ p=n\left(  r_{0}%
,z_{0}\right)  \sin\vartheta_{0}$. As there exist phase restriction for the
variable $z$ then we need in boundary conditions for variables in addition to
the system (\ref{Luch}). The conditions have the form%
\[
\left\{
\begin{array}
[c]{l}%
p\rightarrow p\left(  1-2N_{z}\left(  N_{r}t_{r}/t_{z}+N_{z}\right)  \right)
,\\
z\rightarrow z,
\end{array}
\right.
\]
where $\vec{t}=\left(  t_{r},t_{z}\right)  =\left(  \cos\vartheta
,\sin\vartheta\right)  $ is direction vector of the ray, $\vec{N}=\left(
N_{r},N_{z}\right)  =\left(  \cos\alpha,\sin\alpha\right)  $ is internal
normal to the boundary at the point of ray reflection.

Calculations of amplitudes of the sound pressure deals with the \textit{
matrix} \textit{of variations}%
\[
q=\left\Vert
\begin{array}
[c]{cc}%
q_{11} & q_{12}\\
q_{21} & q_{22}%
\end{array}
\right\Vert \equiv\left\Vert
\begin{array}
[c]{cc}%
\frac{\partial p}{\partial p_{0}} & \frac{\partial p}{\partial z_{0}}\\
\frac{\partial z}{\partial p_{0}} & \frac{\partial z}{\partial z_{0}}%
\end{array}
\right\Vert \
\]
governed by the equation%

\begin{equation}
\frac{d}{dr}q=Kq\label{var}%
\end{equation}
were
\[
K\equiv\left\Vert
\begin{array}
[c]{cc}%
-\frac{\partial^{2}H}{\partial z\partial p} & -\frac{\partial^{2}H}{\partial
z^{2}}\\
\frac{\partial^{2}H}{\partial p^{2}} & \frac{\partial^{2}H}{\partial z\partial
p}%
\end{array}
\right\Vert .
\]
Starting value for matrix $q$ is the identity matrix. Coefficients of the
linear system (\ref{var}) vary through the ray trajectory. When the ray
reaches the waveguide boundary matrix $q$ suffers a transformation of the form%

\[
q\longrightarrow\kappa q.
\]
The transformation matrix%

\begin{gather}
\ \ \kappa=\left\Vert
\begin{array}
[c]{cc}%
\kappa_{11} & \kappa_{12}\\
\kappa_{21} & \kappa_{22}%
\end{array}
\right\Vert ,\nonumber\\%
\begin{array}
[c]{ll}%
\kappa_{11}=-\frac{t_{1r}}{t_{r}}, & \kappa_{12}=\left(  -\mathfrak{K}%
nt_{1r}+N_{z}\left(  \frac{t_{r}^{2}+t_{1r}^{2}}{2t_{r}t_{1r}}-N_{r}%
^{2}\right)  \frac{\partial n}{\partial z}+N_{r}N_{z}\frac{\partial
n}{\partial r}\right)  \frac{2}{t_{N}},\\
\kappa_{21}=0, & \kappa_{22}=-\frac{t_{r}}{t_{1r}},
\end{array}
\label{kkk}%
\end{gather}
where%
\begin{equation}
\vec{t}_{1}=\vec{t}-2\vec{N}N_{t}=\left(  t_{1r},t_{1z}\right)  \label{zercal}%
\end{equation}
is the direction of a ray after reflection of it from the bottom,
$N_{t}=\left\langle \vec{t},\vec{N}\right\rangle .$

In particular case on the free surface $z=0$ and also on the horizontal bottom
$z=\operatorname*{const}$ we have
\[
\ \kappa=\left\Vert
\begin{array}
[c]{cc}%
-1 & 2\frac{n_{z}^{\prime}}{t_{z}}\\
0 & -1
\end{array}
\right\Vert
\]
As $\kappa_{12}\not =0$ when $n_{z}^{\prime}\not \equiv 0$ then horizontal
planes scatter rays in inhomogeneous media.

In homogeneous waveguide with $n=\operatorname*{const}$ we obtain from
(\ref{kkk})%

\[
\ \kappa=\left\Vert
\begin{array}
[c]{cc}%
-\frac{t_{1r}}{t_{r}} & -\mathfrak{K}nt_{1r}t_{r}\frac{2}{N_{t}}\\
0 & -\frac{t_{r}}{t_{1r}}%
\end{array}
\right\Vert
\]

\begin{figure}[ptb]
\centering\subfloat[]{\label{fig:kap1}\includegraphics[width=0.33\textwidth]{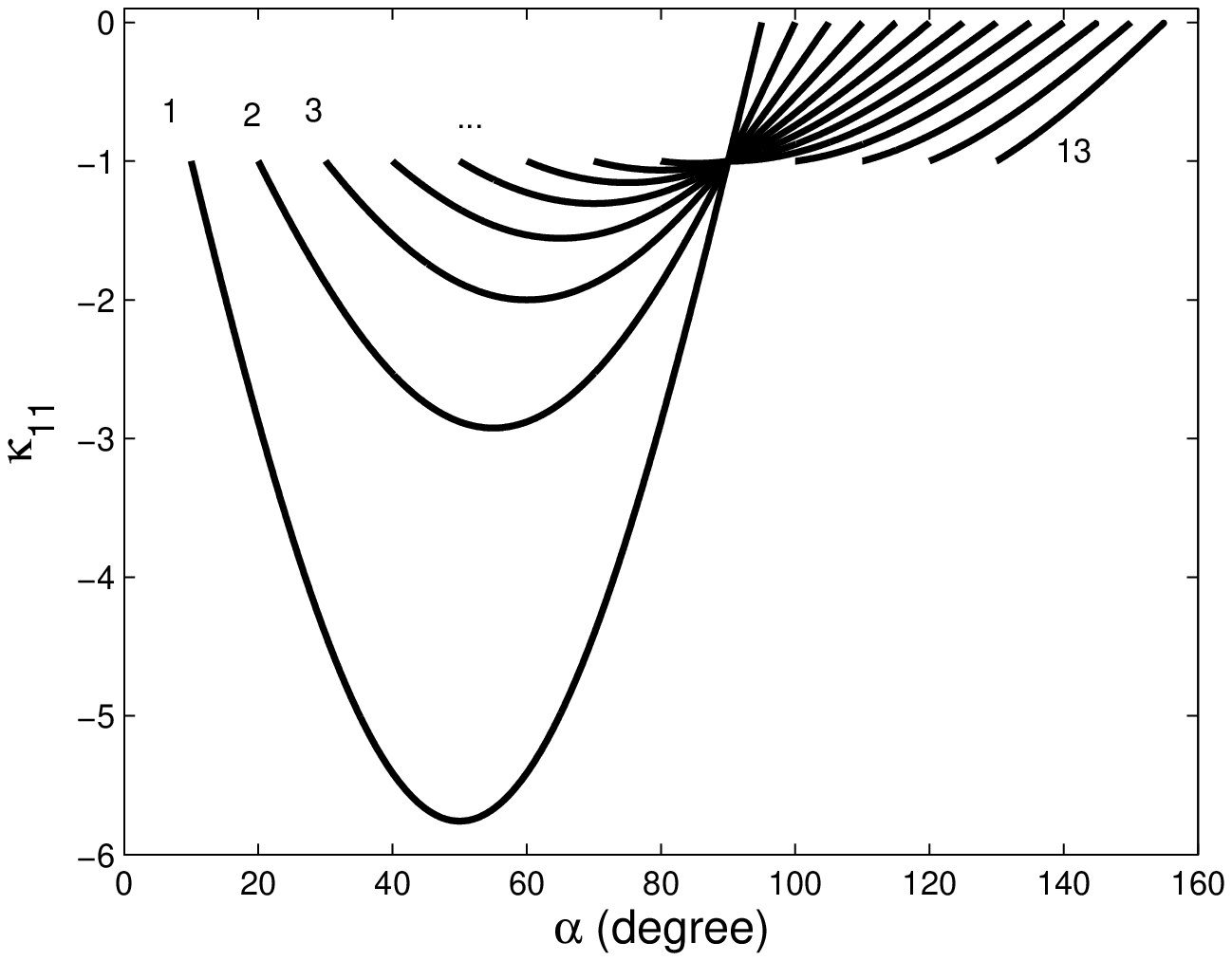}}\subfloat[]{\label{fig:kap2}\includegraphics[width=0.33\textwidth]{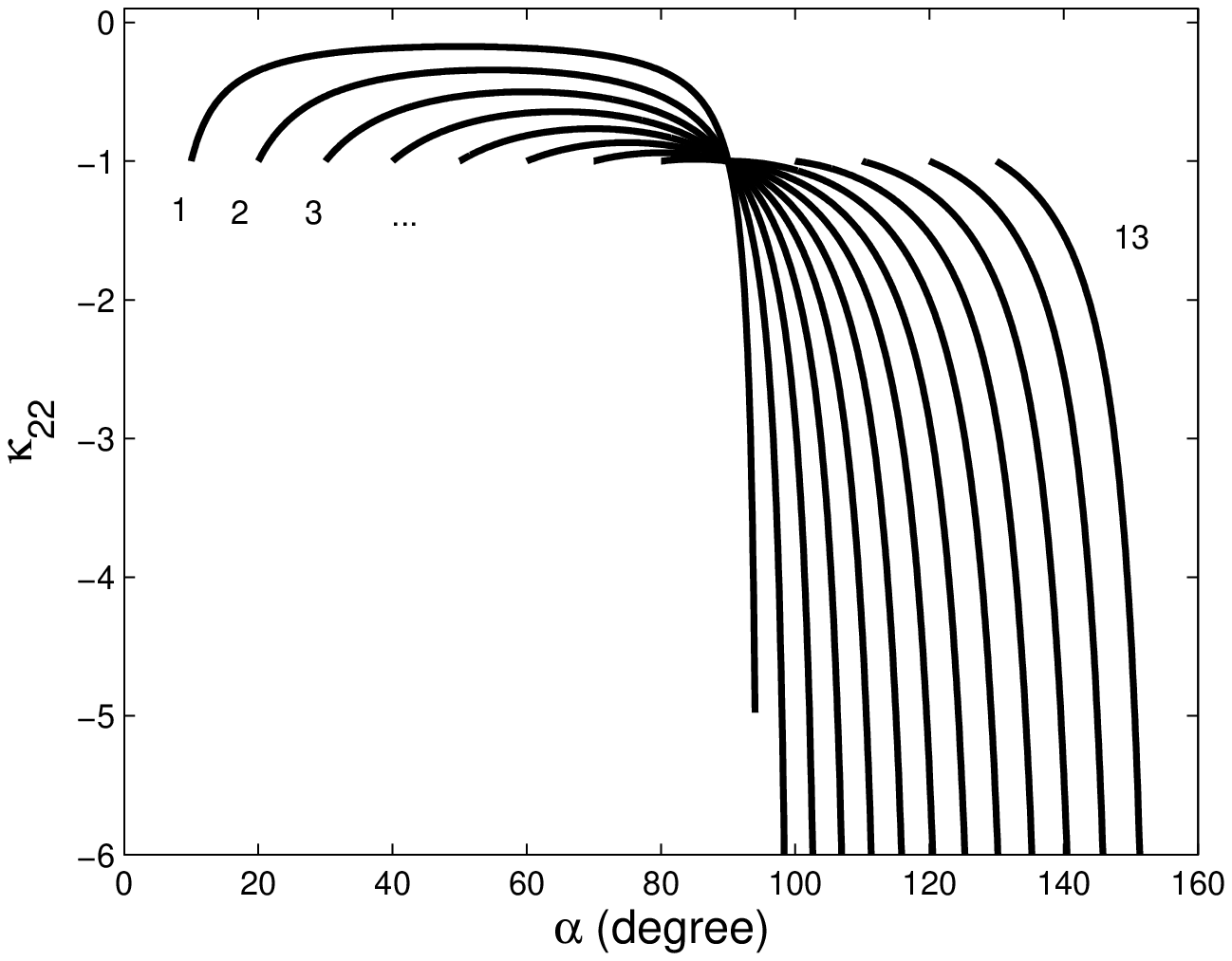}}\subfloat[]{\label{fig:kap2}\includegraphics[width=0.33\textwidth]{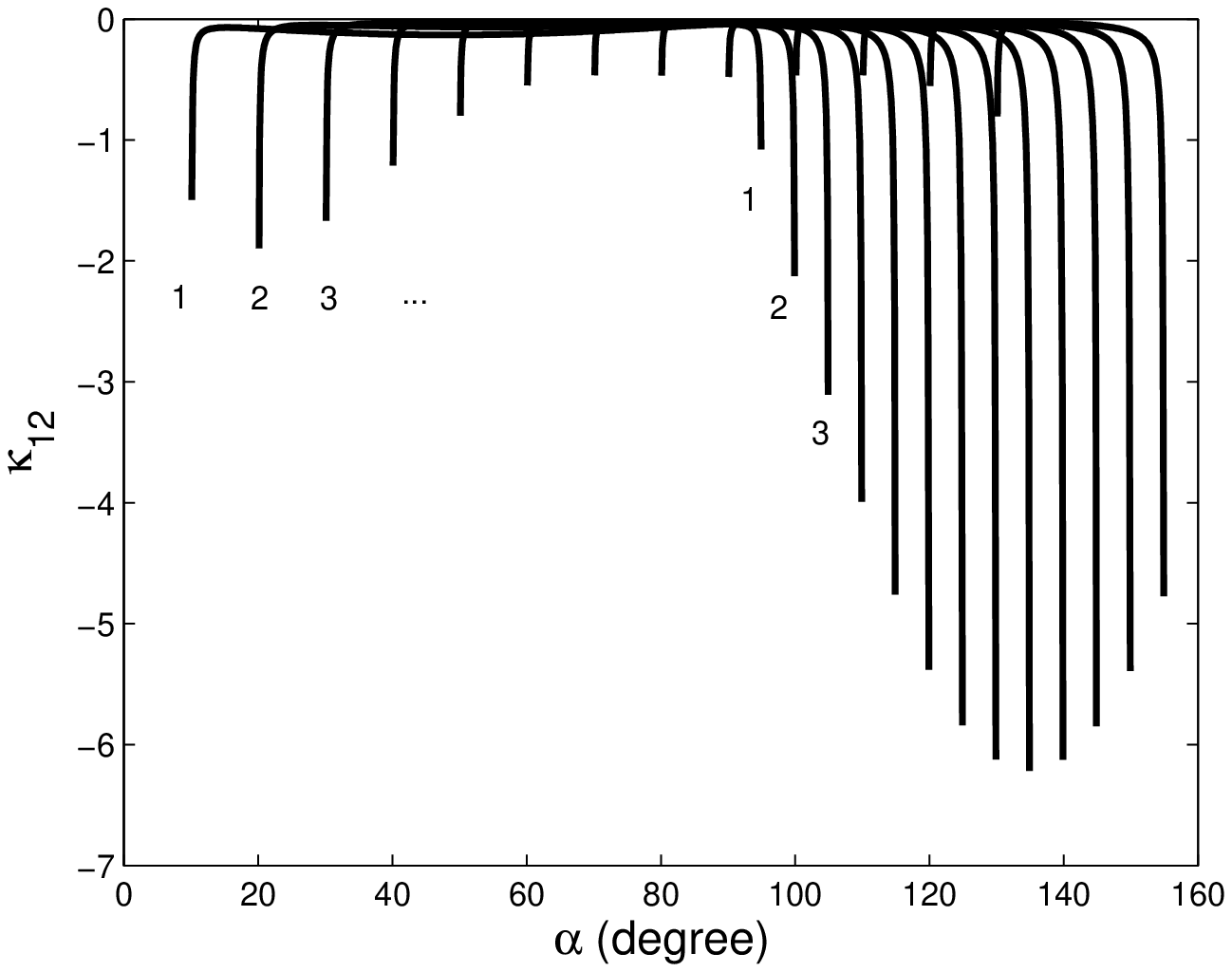}}\caption{Coefficients
$\kappa_{11}$ (a), $\kappa_{22}$ (b) and $\kappa_{12}$ (c) for fixed ray
gracing angles $\vartheta$ versus  angle $\alpha$; the curve number $k$
corresponds to the angle $\vartheta=-90^{\circ}+10^{\circ}k$; the curve at the
point of reflection $\mathfrak{K}=0.02$, derivatives $n_{z}^{\prime}=0.01$,
$n_{r}^{\prime}=0$.}%
\label{fig:kappas}%
\end{figure}

To prove formula (\ref{kkk}) let consider a narrow beam of rays, bounded by
the rays with similar pulses $p$ and $\tilde{p}$. Let the incident points of
the rays be $\left(  r,z\right)  $ and $\left(  \tilde{r}=r+\delta r,\tilde
{z}=z+\delta z\right)  $, direction vectors $\vec{t}$ and $\widetilde{\vec{t}%
}$, normals $\vec{N}$ and $\widetilde{\vec{N}}$, respectively.

\begin{figure}[ptb]
\centering\subfloat[]{\label{fig:refa}\includegraphics[width=0.5\textwidth]{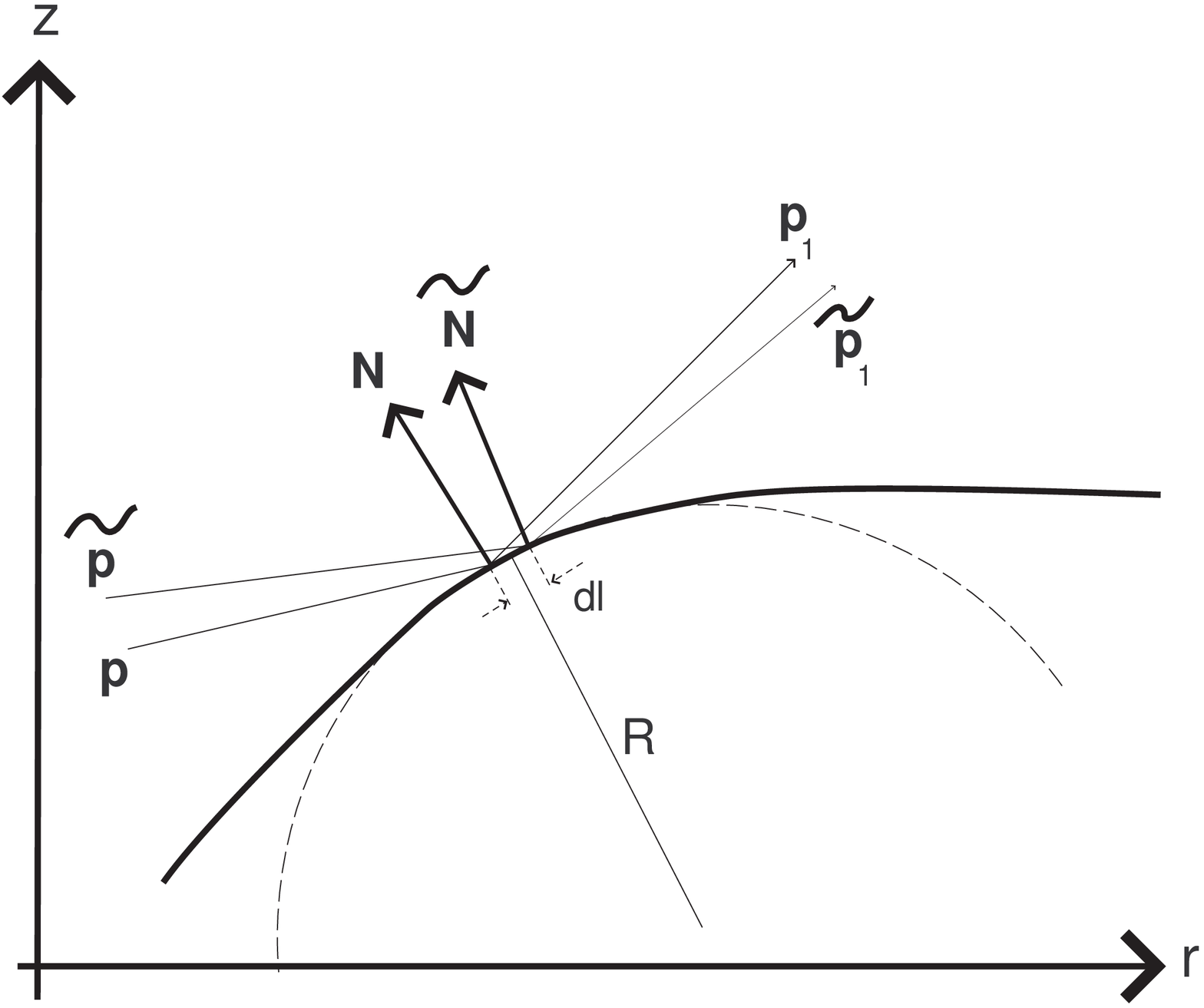}}
\subfloat[]{\label{fig:refb}\includegraphics[width=0.5\textwidth]{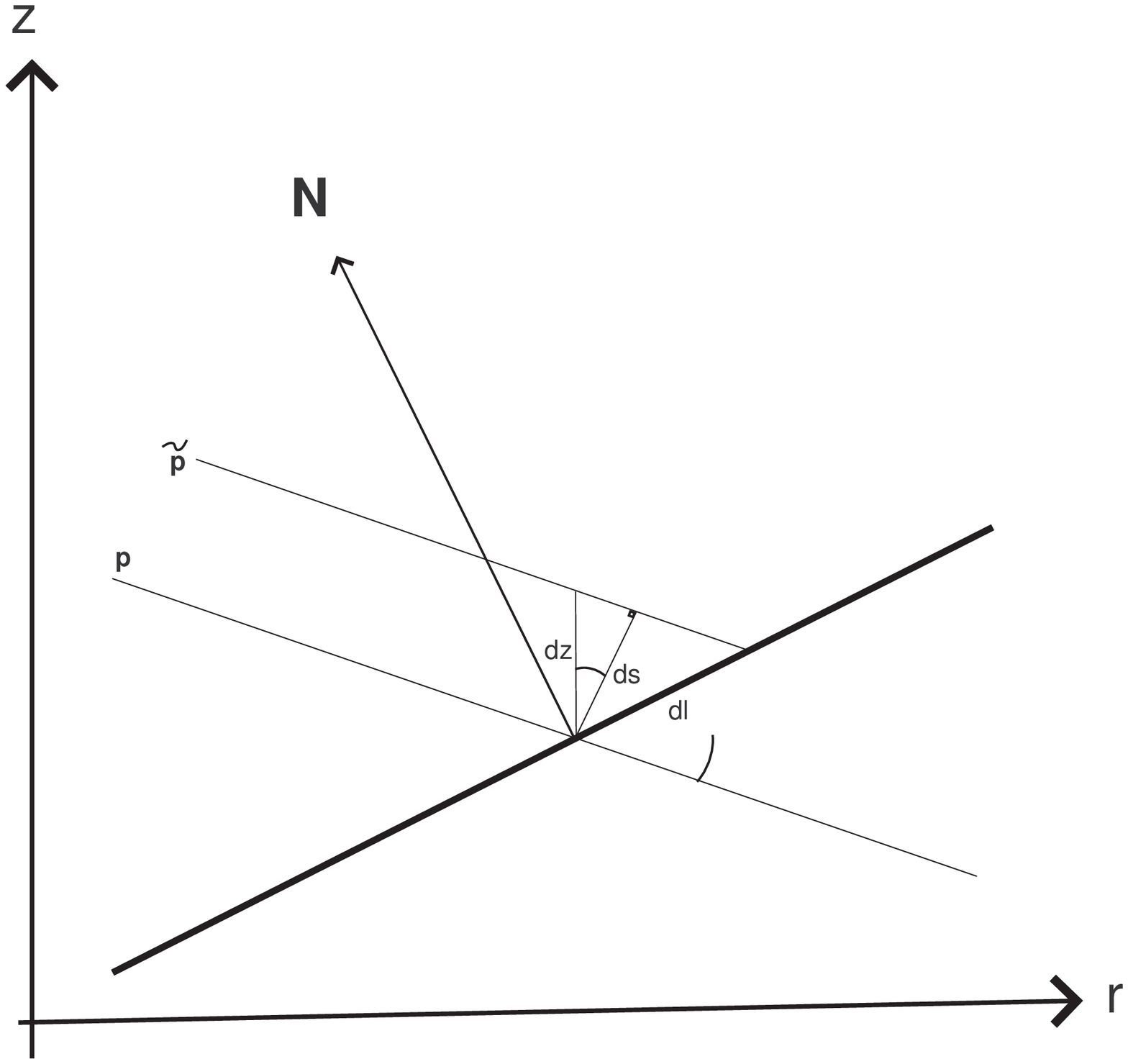}}\caption{Scheme
reflection of a narrow beam of rays from the curved surface with normal vector
$\vec{N}$ and the radius of curvature $R.$}%
\end{figure}

From geometrical reason (Fig. \ref{fig:refb}) we receive%
\begin{gather*}
\delta r\simeq\frac{t_{r}N_{z}}{N_{t}}\Delta z,\\
\delta z\simeq-\frac{t_{r}N_{r}}{N_{t}}\Delta z.
\end{gather*}

First, we find the laws of reflection of the beam, neglecting the change of
the normal vector at the site of incidence on the surface, that is, assuming a
flat surface. For the increment of $Δ z=z(r)-z(r)$ we have in this approximation%

\begin{gather}
z_{1}\left(  r+\delta r\right)  =z_{1}\left(  r\right)  +\frac{d}{dr}%
z_{1}\delta r=z\left(  r\right)  +\frac{t_{1z}}{t_{1r}}\delta r,\nonumber\\
\tilde{z}_{1}\left(  r+\delta r\right)  =\tilde{z}\left(  r\right)  +\frac
{d}{dr}\tilde{z}\delta r=\tilde{z}+\frac{\tilde{t}_{z}}{\tilde{t}_{r}}\delta
r,\nonumber\\
\Delta z_{1}\equiv z_{1}\left(  r+\delta r\right)  -\tilde{z}\left(  r+\delta
r\right)  =\Delta z+\left(  \frac{t_{1z}}{t_{1r}}-\frac{\tilde{t}_{z}}%
{\tilde{t}_{r}}\right)  \delta r\simeq\nonumber\\
\simeq\left(  1+\frac{t_{r}N_{z}}{N_{t}}\left(  \frac{t_{1z}}{t_{1r}}%
-\frac{t_{z}}{t_{r}}\right)  \right)  \Delta z=-\frac{t_{r}}{t_{1r}}\Delta
z.\label{deltaz}%
\end{gather}
For $\Delta p=p\left(  r\right)  -\tilde{p}\left(  r\right)  $ we have
analogously:%
\begin{gather}
p_{1}\left(  r+\delta r\right)  =p_{1}\left(  r\right)  +\frac{d}{dr}%
p_{1}\delta r=\nonumber\\
=p\left(  r\right)  -2N_{z}\left(  N_{r}\sqrt{n^{2}-p^{2}\left(  r\right)
}+p\left(  r\right)  N_{z}\right)  +\frac{n_{z}^{\prime}}{t_{1r}}\delta
r,\nonumber\\
\tilde{p}_{1}\left(  r+\delta r\right)  =\tilde{p}\left(  r+\delta r\right)
-2N_{z}\left(  N_{r}\sqrt{n^{2}\left(  r+\delta r,z+\delta z\right)
-\tilde{p}^{2}\left(  r+\delta r\right)  }+\tilde{p}\left(  r+\delta r\right)
N_{z}\right)  ,\nonumber\\
\tilde{p}\left(  r+\delta r\right)  =\tilde{p}\left(  r\right)  +\frac{d}%
{dr}\tilde{p}\delta r=\tilde{p}\left(  r\right)  +\frac{n_{z}^{\prime}}{t_{r}%
}\delta r,\nonumber\\
\Delta p_{1}\equiv p_{1}\left(  r+\delta r\right)  -\tilde{p}\left(  r+\delta
r\right)  =\nonumber\\
\simeq\Delta p-2N_{z}N_{r}\left(  \sqrt{n^{2}-p^{2}\left(  r\right)  }%
-\sqrt{n^{2}\left(  r+\delta r,z+\delta z\right)  -\tilde{p}^{2}\left(
r+\delta r\right)  }\right)  -\nonumber\\
-2N_{z}\left(  p\left(  r\right)  -\tilde{p}\left(  r+\delta r\right)
\right)  N_{z}+\left(  \frac{d}{dr}p_{1}-\frac{d}{dr}\tilde{p}\right)  \delta
r=\nonumber\\
=\Delta p\left(  1-2N_{z}^{2}+2N_{z}N_{r}\frac{t_{z}}{t_{r}}\right)  +\left(
\frac{d}{dr}p_{1}-\frac{d}{dr}\tilde{p}\left(  1-2N_{z}^{2}+2N_{z}N_{r}%
\frac{t_{z}}{t_{r}}\right)  \right)  \delta r+\nonumber\\
+2N_{z}N_{r}\frac{n}{\sqrt{n^{2}-p^{2}\left(  r\right)  }}\left(
n_{r}^{\prime}\delta r+n_{z}^{\prime}\delta z\right)  =\nonumber\\
=-\frac{t_{1r}}{t_{r}}\Delta p+\left[  \left(  \frac{t_{r}}{t_{1r}}%
+\frac{t_{1r}}{t_{r}}-2N_{r}^{2}\right)  n_{z}^{\prime}+2N_{r}N_{z}%
n_{r}^{\prime}\right]  \frac{N_{z}}{N_{t}}\Delta z.\label{deltap1}%
\end{gather}
Here we have used the identity%
\begin{gather*}
1-2N_{z}^{2}+2N_{z}N_{r}\frac{t_{z}}{t_{r}}\equiv-\frac{t_{1r}}{t_{r}},\\
1+\frac{t_{r}N_{z}}{N_{t}}\left(  \frac{t_{1z}}{t_{1r}}-\frac{t_{z}}{t_{r}%
}\right)  \equiv-\frac{t_{r}}{t_{1r}}.
\end{gather*}

Now take into account the effect of the curvature of the surface at the point
of reflection, neglecting variations in $\vec{t}$ and $n$. Let the radius of
curvature of the bottom surface is equal to $R$, and she curvature
$\mathfrak{K}=1/R$. 

We have%

\begin{align*}
dN_{z} &  =d\sin\alpha=N_{r}d\alpha,\\
dN_{r} &  =d\cos\alpha=-N_{z}d\alpha.
\end{align*}
Differentiating (\ref{zercal}), we obtain%

\begin{gather*}
-dt_{1z}/2=dN_{z}\left\langle \vec{t},\vec{N}\right\rangle +N_{z}\left\langle
\vec{t},d\vec{N}\right\rangle =\\
=dN_{z}\left(  \left\langle \vec{t},\vec{N}\right\rangle +t_{z}N_{z}\right)
+N_{z}t_{r}dN_{r}=\\
=\left(  N_{r}\left(  2t_{z}N_{z}+t_{r}N_{r}\right)  -N_{z}^{2}t_{r}\right)
d\alpha=-t_{1r}d\alpha.
\end{gather*}
Then (see Fig. \ref{fig:refb})%
\[
d\alpha\simeq-\mathfrak{K}dl=-\mathfrak{K}\frac{\delta r}{N_{z}}%
=-\mathfrak{K}\frac{t_{r}}{N_{t}}\Delta z,
\]
so when multiplied by $n$ we get%

\begin{equation}
\Delta p_{1}=2nt_{1r}d\alpha\simeq-2\mathfrak{K}n\frac{t_{1r}t_{r}}{N_{t}%
}\Delta z.\label{deltap2}%
\end{equation}

After summation of (\ref{deltap1}), (\ref{deltap2}) and passing to the limit
as $\Delta p_{0}\rightarrow0$ we obtain (\ref{kkk}).

I wish to thank A.L. Virovlyasnky for useful discussions.


\begin{thebibliography}{9}                                                                                                %
\bibitem {Viro}A.L. Virovlyansky. Ray theory of long-range sound propagation
in the ocean (in Russian). Nizhny Novgorod: IAP\ RAS, 2006.
\end{thebibliography}
\end{document}